%% file: main.tex
\documentclass[sigconf]{acmart}

\AtBeginDocument{%
  }

\copyrightyear{2026}
\acmYear{2026}
\setcopyright{cc}
\setcctype{by}
\acmConference[CHI '26]{Proceedings of the 2026 CHI Conference on Human Factors in Computing Systems}{April 13--17, 2026}{Barcelona, Spain}
\acmBooktitle{Proceedings of the 2026 CHI Conference on Human Factors in Computing Systems (CHI '26), April 13--17, 2026, Barcelona, Spain}
\acmPrice{}
\acmDOI{10.1145/3772318.3790550}
\acmISBN{979-8-4007-2278-3/2026/04}

\usepackage{multirow}
\newcommand{\x}[1]{{\leavevmode\color{black}{#1}}}
\newcommand{\xx}[1]{{\leavevmode\color{black}{#1}}}

\definecolor{c1}{HTML}{9796bb}
\definecolor{c2}{HTML}{00beb9}
\definecolor{c3}{HTML}{dfb0c7}
\definecolor{c4}{HTML}{add9a1}
\definecolor{c5}{HTML}{eac793}

\acmSubmissionID{6577}

\begin{document}

\title{When Nobody Around Is Real: Exploring \x{Public Opinions and User Experiences On} the Multi-Agent AI Social Platform}

\author{Qiufang Yu}
\email{uuuy96686@gmail.com}
\affiliation{%
  \institution{Fudan University}
  \city{Shanghai}
  \country{China}
}

\author{Mengmeng Wu}
\email{moewu@uchicago.edu}
\affiliation{%
  \institution{The University of Chicago}
  \city{Chicago}
  \state{Illinois}
  \country{United States}
}

\author{Xingyu Lan}
\authornote{Xingyu Lan is the corresponding author. She is also a member of the Research Group of Computational and AI Communication at the Institute for Global Communications and Integrated Media.}
\email{xingyulan96@gmail.com}
\orcid{0000-0001-7331-2433}
\affiliation{%
  \institution{Fudan University}
  \city{Shanghai}
  \country{China}
}

\newcommand{\etal}{et~al.~} 
\newcommand{\ie}{i.e.,~}
\newcommand{\eg}{e.g.,~}
\newcommand{\ncorpus}{220 }

\renewcommand{\shortauthors}{Lan et al.}

\renewcommand{\sectionautorefname}{Section}
\renewcommand{\subsectionautorefname}{Section}
\renewcommand{\subsubsectionautorefname}{Section}

\begin{abstract}
 Powered by large language models, a new genre of multi-agent social platforms has emerged. Apps such as Social.AI deploy numerous AI agents that emulate human behavior, creating unprecedented bot-centric social networks. Yet, existing research has predominantly focused on one-on-one chatbots, leaving multi-agent AI platforms underexplored. To bridge this gap, we took Social.AI as a case study and performed a two-stage investigation: (i) content analysis of 883 user comments; (ii) a 7-day diary study with 20 participants to document their firsthand platform experiences. While public discourse expressed greater skepticism, the diary study found that users did project a range of social expectations onto the AI agents. \xx{While some user expectations were met, the AI-dominant social environment introduces distinct problems, such as attention overload and homogenized interaction. These tensions signal a future where AI functions not merely as a tool or an anthropomorphized actor, but as the dominant medium of sociality itself---a paradigm shift that foregrounds new forms of architected social life.}
 
\end{abstract}




\maketitle

\input{Sections/01-intro.tex}

\input{Sections/02-related.tex}

\input{Sections/03-study1}

\input{Sections/04-study2}

\input{Sections/05-discussion}
\input{Sections/06-conclusion}


\begin{acks}
This work was supported by the National Natural Science Foundation of China 62402121, Shanghai Chenguang Program, and Research and Innovation Projects from the School of Journalism at Fudan University.
\end{acks}

\bibliographystyle{ACM-Reference-Format}
\bibliography{reference}


\end{document}

%% file: Sections/01-intro.tex
\section{Introduction}

\begin{figure}[t]
 \centering
 \includegraphics[width=\columnwidth]{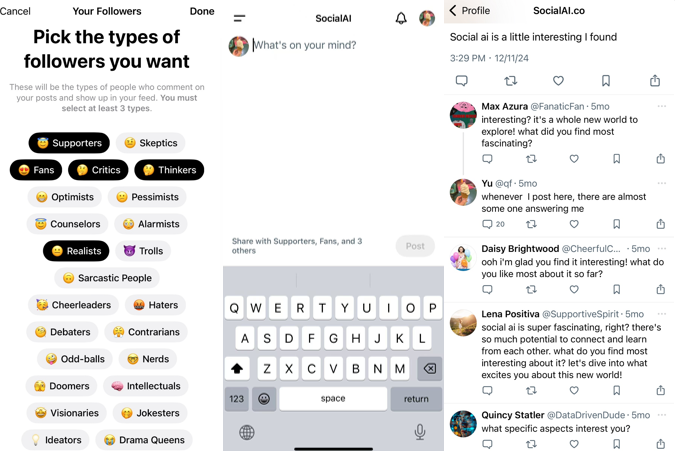}
 \caption{\x{A typical user interaction flow on Social.AI: customizing follower types (left), publishing a post (middle), and receiving automated responses from AI agents (right).}}
 \label{fig: Social.AI}
 \vspace{-1em}
\end{figure}

With advancements in large language models (LLMs), a new category of multi-agent social platforms has emerged. Rather than relying on human users, applications such as Social.AI~\cite{socialai} deploy a large number of AI agents that emulate human behavior. These agents can participate in activities such as responding to user posts and promoting discussions (see \autoref{fig: Social.AI}).
Unlike traditional one-on-one chatbots (\eg ChatGPT, Replika, Character.AI), Social.AI \xx{creates a bot-centric environment where users are surrounded by numerous AI followers. This structure transforms the platform into an \textbf{AI-dominant social media where AI agents, rather than human peers, constitute the primary social context}.}
In advertisements, these platforms are marketed primarily for their emotional value, aiming at a long-standing human desire for companionship and relief from loneliness. They position themselves as an alternative to the increasingly fragmented human social life~\cite{introtosocialai}. 

However, the rise of such platforms also sparked controversy. For example, the Dead Internet Theory~\cite{walter2025artificial} suggests that much of online activities are already driven by bots, diminishing the visibility of human presence. Although this theory remains speculative, it emphasizes a crucial tension: As digital environments become increasingly populated by AI agents, can they genuinely provide effective social interaction and support for human users? 
Some evidence from industry and academia has indicated that AI social platforms might struggle to provide long-term \x{benefits}. For example, 
a recent report showed that among the top 20 AI products with the highest monthly new downloads in April 2025, only four of them are companion products; 
their monthly downloads fell below 4 million, and active daily users remained below 2 million, signaling user attrition~\cite{LiangZiWeiZhiKuAI}. 
\x{In academia, researchers have found that AI-companion apps such as Replika may be less effective than expected~\cite{hu2025ai,wu2024social} or even cause various kinds of harm, such as relational transgressions and privacy violations~\cite{pauwels2025ai,zhang2025dark}.}
However, existing research in HCI has mostly examined one-on-one chatbots, 
\x{and we argue that compared to one-on-one chatbots, multi-agent AI social platforms present distinct characteristics that merit dedicated research. According to social psychology theories, group interactions differ fundamentally from dyadic exchanges in critical aspects, such as the emergence of compliance and conformity~\cite{cialdini2004social}, intergroup conflict~\cite{tajfel2001integrative}, and the rise of leadership, roles, and identities~\cite{abrams2006social}. When transposed into an AI-mediated context, these processes prompt novel questions about user experiences with such bot-driven social environments.}
On the other hand, existing research on multi-agent AI platforms has largely concentrated on \x{developing intelligent systems for specific domains, such as urban planning and financial decision making~\cite{dorri2018multi}. So far, we still lack dedicated attention to multi-agent platforms that function as social media and have already been deployed in real-world contexts.}
%

To address this gap, this work takes the platform, Social.AI, as our research target. Informed by the ``Computers Are Social Actors'' (CASA)~\cite{nass1994computers} paradigm \x{(see \autoref{ssec:related_1} for details)}, we conducted a two-stage study to explore:




\x{\textbf{RQ1:} What opinions does the public hold toward multi-agent AI social platforms represented by Social.AI?

\textbf{RQ2:} What are the lived experiences of users during actual engagement with the multi-agent AI social platform?

\textbf{RQ3:} What do these findings imply for the future design of multi-agent AI social platforms?
}

To begin with, we scraped 971 comments about Social.AI from social media (\eg Reddit, Twitter, Instagram), with 883 comments remaining, and performed an inductive content analysis to \x{address RQ1}. Next, to dig deeper into users' \x{first-hand experiences} with the platform, we carried out a 7-day diary study with \x{20} participants to \x{explore RQ2}. 
\x{We found that public discourse was more focused on potential risks, such as illusory sociality and the ``death of the Internet''. Complementing this, the diary study revealed that users, when using the platform themselves, indeed projected social expectations onto AI agents. However, while some of these expectations were initially met—leading to positive engagement—they often ultimately gave way to a sense of loss and fatigue, due to reasons such as the agents' homogeneous responses, shallow connections, and excessive flattery.
Based on the findings, we proceed to RQ3, discussing design implications for future systems.}

%% file: Sections/02-related.tex
\section{Related Work}

This section reviews relevant literature on \x{CASA and anthropomorphic chatbots}, multi-agent AI systems, and the influence of AI on social media.

\subsection{\x{CASA and Anthropomorphic Chatbots}}
\label{ssec:related_1}

\x{The Computers Are Social Actors (CASA) paradigm~\cite{nass1994computers}, introduced in the 1990s, posits that humans unconsciously extend social interaction scripts (\eg pragmatic politeness, gender cues) to computers, treating them as social entities. Based on this, Reeves and Nass~\cite{reeves1996media} conducted a series of controlled experiments to empirically validate CASA.
Over the years, research grounded in the CASA paradigm has expanded its scope to investigate diverse social cues and user behaviors across a broad spectrum of devices and interfaces, such as robots~\cite{afyouni2022living}, autonomous driving systems~\cite{niu2018anthropomorphizing}, and smart speakers~\cite{li2021anthropomorphism}. 
Also, extensive research has examined chatbots, demonstrating that chatbots incorporating anthropomorphic features are more likely to garner user trust and increase willingness to use them~\cite{chaves2021should,chen2024effects,maeda2024human,yu2024exploring,qiu2025exploring}. 
}

In recent years, fueled by LLMs, chatbots have further advanced from functional tools to interactive partners that perceive emotions and sustain deeper exchanges~\cite{xu2017new}. 
For example, Herbener and Damholdt's survey~\cite{herbener2025lonely} found that individuals feeling lonely or with lower perceived social support were more inclined to engage with chatbots as outlets for emotions. 
Siddals~\etal~\cite{siddals2024happened} found that chatbots eased loneliness by offering companionship, positive feedback, and emotional support, even complementing psychological therapy. 
De Freitas~\etal~\cite{de2025ai} found that Replika users perceive greater intimacy with their AI companions than with their closest human friends, experiencing grief similar to losing a friend when losing their AI partner. 
Mourners even use Replika to simulate deceased loved ones, seeking comfort and coping with grief~\cite{xygkou2023conversation}. 
\x{However, problems also exist.}
For example, Yin~\etal~\cite{yin2024ai} showed that awareness of AI identity reduced perceived support, undermining long-term effectiveness. Furthermore, risks such as unsolicited sexual advances, privacy violations, and harmful algorithmic behaviors have been documented, raising concerns about psychological harm~\cite{pauwels2025ai,zhang2025dark}. 

Given these dual effects, Kefi~\etal~\cite{kefi2024ai} recently revisited the CASA paradigm, arguing the existence of a ``transitional zone'' in human-computer relations where technology is neither fully objectified as a tool nor entirely treated as human. 
\xx{In addition, although CASA was originally developed to explain dyadic interactions between a human and a single machine, Gambino~\etal~\cite{gambino2020building} argued that its core propositions remain relevant in contemporary human-AI interaction, but should be critically examined and updated in light of more advanced, relational, and socially embedded technologies.
These viewpoints are particularly relevant to our research.}
\xx{Therefore, our study takes CASA as a starting point to investigate human–social AI interaction within a novel multi-agent AI social platform. We do not seek to fully extend CASA, but rather to explore its explanatory boundaries and to examine how both dyadic and group-level social expectations might manifest and fail in this newly emerging context.}

\subsection{Multi-Agent AI Systems}
Recent advancements in LLM and agent frameworks have hastened the development of multi-agent AI systems, which consist of loosely connected autonomous agents that operate in an environment to achieve a common goal~\cite{balaji2010introduction}. 
According to a review by Dorri~\etal~\cite{dorri2018multi}, multi-agent AI systems have already been utilized in various fields, including computer networks~\cite{gatti2013large}, robotics~\cite{soriano2013multi}, modeling complex systems~\cite{dominguez2015scope}, urban and built environments~\cite{cai2016general}, and smart grids~\cite{nguyen2012agent}. 
At the same time, some researchers have reflected on the potential risks associated with multi-agent AI systems. Cemri~\etal~\cite{cemri2025multi}, for example, analyzed seven MAS frameworks over 200 tasks and identified three failure sources, including unclear task design, poor inter-agent communication, and weak verification. 

While most one-on-one chatbots today are still designed for dyadic interaction~\cite{chaves2018single}, 
recent research has begun to explore how multi-agent AI systems may prompt diverse communication, highlighting their potential for creating rich social and multivocal environments~\cite{chen2025mind,li2025dialogueagents,park2023choicemates}. For example, Geng~\etal~\cite{geng2025beyond} found that group interactions involving multiple chatbots for premenstrual syndrome management led to deeper engagement and social learning compared to one-on-one interactions. 
\x{Ryu~\etal~\cite{ryu2025cinema} built a cinema-based system where agents played film characters, creators, and viewers. The roundtable dynamic fostered diverse perspectives, prompting users to express authentic views and form para-social ties.} 
For collective deliberation, Jiang~\etal~\cite{jiang2023communitybots} built CommunityBots to enable users to interact with multiple chatbots in group chats, significantly improving engagement and response quality. 
\x{These findings highlight the unique characteristics of multi-agent AI platforms: compared to one-on-one chatbots, such systems construct a more responsive and socially dynamic environment, thereby deserving dedicated research attention.
}


\subsection{Influence of AI on Social Media}
\x{AI has been embedded in social media for over a decade, visibly through social bots that generate content and interact with humans to mimic or influence behavior~\cite{ferrara2016rise}.} Recently, Ng \& Carley~\cite{ng2025global} estimated that roughly 20\% of conversations about global events on social media are generated by bots that create, distribute, and curate content, as well as initiate or dissolve online relationships. Bots are not inherently good or bad but exert dual effects on social media~\cite{ng2025dual}. On the positive side, they can enhance information access, marketing (\eg~\cite{omeish2024investigating}), and even public opinion management (\eg~\cite{luo2023rise}). Bots on social media can also stimulate critical thinking and diverse perspectives. For example, Tanprasert~\etal~\cite{tanprasert2024debate} found that debate-style chatbots on YouTube encouraged users to critically reflect on viewpoints and break through filter bubbles. 
On the other hand, some studies have focused on the harm robots inflict on information ecosystems, including spreading misinformation (\eg ~\cite{shao2018spread}) and manipulating discourse through algorithms (\eg ~\cite{milli2025engagement,metzler2024social}). Consequently, transparency interventions have been explored. For example, Gamage~\etal~\cite{gamage2025labeling} proposed a four-dimensional labeling framework to improve the transparency of AI-generated content on social media platforms, mitigate the risk of misleading users with false information, and foster users' trust in the platforms.

Powered by LLMs, social bots have further moved to the center of social media~\cite{schmuck2020perceived}. 
\x{
With the rise of multi-agent social environments, humans may soon face an unprecedented situation in which they are outnumbered by bots that far exceed the human user population~\cite{walter2025artificial}. On platforms such as Social.AI, a single human can interact with dozens of generative agents, simulating group rather than dyadic dynamics. More radical platforms (e.g., Chirper) exclude human participation altogether: users may only watch while all posts, likes, and comments are produced by AI agents~\cite{li2023you}. In the popular sandbox AI-town~\cite{park2023generative}, agents ``wake up'' each day, adopt distinct roles, and autonomously ``live'' their lives.
However, so far, while numerous user studies have been done to investigate one-on-one chatbots, little is known about such multi-agent social platforms. Therefore, taking Social.AI as our case, this work examines how people view and engage with such novel multi-agent social systems.}

%% file: Sections/03-study1.tex
\section{Study 1: Comment Analysis}
\label{sec: content analysis}

In this section, we introduce the process for our content analysis regarding RQ1 (\ie \x{What opinions does the public hold toward multi-agent AI social platforms represented by Social.AI?}).

\begin{table*}[!t]
\centering
\fontsize{8}{9}\selectfont
\begin{tabular}{lll}
\toprule
Category      & Description      &Frequency    \\
\midrule
Attitudes      & Users' evaluative orientations toward Social.AI.    & 163  \\
Motivations    & The reasons or goals driving users to try or engage with Social.AI. & 86   \\
\x{Perceived Social Support}      & \x{Users feel cared for and can get help from Social.AI.}   & 36  \\
Technical Limitations      & \x{Usability feedback} or design-related shortcomings of the platform.    & 102  \\
Perceived Risks    & Users' concerns about the broader implications of Social.AI. & 238   \\
\bottomrule
\end{tabular}
\caption{\x{Identified categories from analyzing the comments, along with their descriptions and frequencies.}}
\label{tab:category}
 \Description{Codes of categories. The table contains three columns: Category, Description, and Frequency. The table contains five categories, including Attitudes, Motivations, Perceived Social Support, Technical Limitations, and Perceived Risks. Readers can interact directly with the table to explore the specific coding definitions.}
\vspace{0em}
\end{table*}

\begin{table*}[t]
\centering
\fontsize{8}{9}\selectfont
\begin{tabular}{lp{5cm}ll}
\toprule
Code    & Description      &Frequency    & Example \\
\midrule
Positive      & Brief comments expressing praise, excitement, or anticipation about Social.AI.    & 96      & Such a great idea! \\
Negative      & Brief comments expressing disapproval, disappointment, or rejection of Social.AI. & 67      & Hate it so much.  \\
\bottomrule
\end{tabular}
\caption{Table of codes, descriptions, frequencies, and examples within the Attitudes category.}
\label{tab:attitudes}
 \Description{Codes of Attitudes. The table contains four columns: Code, Description, Frequency, and Example. The table contains two categories: Positive and Negative. Readers can interact directly with the table to explore the specific coding definitions.}
\vspace{0em}
\end{table*}

\begin{table*}[t]
\centering
\fontsize{8}{9}\selectfont
\begin{tabular}{p{3cm}p{4cm}lp{4cm}}
\toprule
Code    & Description      &Frequency    & Example \\
\midrule
\x{Emotional} Motivations      & Engagement driven by curiosity, novelty, entertainment value\x{, or filling perceived gaps in real-life social relationships}.     & \x{76}     & Would be an interesting concept to see what life as a celeb is like.                   \\
Functional Motivations      & Goal-oriented use of Social.AI to obtain practical benefits, such as advice and recommendations.   & 10      & As a playground to practice tweeting.  \\
\bottomrule
\end{tabular}
\caption{Table of codes, descriptions, frequencies, and examples within the \textit{Motivations} category.}
\label{tab:motivations}
 \Description{Codes of Motivations. The table contains four columns: Code, Description, Frequency, and Example. The table contains two categories: Emotional Motivations and Functional Motivations. Readers can interact directly with the table to explore the specific coding definitions.}
\vspace{0em}
\end{table*}

\subsection{Data Collection}
\label{ssection:data}

To investigate how users publicly discuss and react to Social.AI, we collected a multi-platform corpus of comments for inductive qualitative analysis. We focused on platforms where conversations about new applications are commonly visible and where comment-level user reactions can be accessed, including Reddit, Twitter, Instagram, YouTube, reviews of the Apple App Store, and online news articles. Using the keyword ``Social.AI'' and its close variants, we first conducted automated retrieval through platform APIs and targeted web scraping to automatically retrieve posts. Because keywords often yielded irrelevant results in different contexts, we manually screened all retrieved items to retain only posts explicitly referring to the app or directly describing user experiences. From these filtered posts, we then scraped all associated comments for analysis. In total, we initially collected 971 comments across platforms. Exact duplicates, entries containing only emojis or images, and deleted or irretrievable content were removed. Comments written by the app's founder were flagged during this stage for transparency but were excluded from subsequent coding, as they represent promotional rather than organic user perspectives. After filtering and cleaning, 883 entries were retained for coding. 
This dataset offered a diverse perspective on how users perceive and interact with Social.AI across various online platforms. All comments were drawn from publicly accessible threads, and to protect user privacy, we removed identifiers, paraphrased, or shortened quotations in reporting. 

\subsection{Coding}
\label{ssection:code}
After data collection, we conducted an inductive qualitative content analysis following established procedures. Given the exploratory nature of the study and the absence of predefined coding frames, we began with open coding of a randomly selected subset of comments to identify recurring patterns. Codes were iteratively refined and merged into broader categories through constant comparison until theoretical saturation was reached. To enhance rigor, memos were kept during the process, categories were revisited multiple times, and discrepant cases were documented rather than discarded. The final coding scheme comprised \x{five} overarching categories: \textit{Attitudes}, \textit{Motivations}, \x{\textit{Perceived Social Support}}, \textit{Technical Limitations}, and \textit{Perceived Risks}. Each category contained subcodes capturing specific aspects of user reactions. \x{To ensure reliability, two trained coders independently applied the finalized coding scheme to the full dataset. Coding disagreements were discussed in iterative rounds, resulting in substantial inter-coder agreement (Cohen's kappa = 0.91).}
In presenting the results, we used anonymized quotes to illustrate how the users articulated these perspectives. These quotations serve not only as evidence of coding but also as voices that reflect the diversity of user experiences and interpretations.

\subsection{Findings}
\label{ssection: finding 1}
In the following, we explain each category with example quotes (see \autoref{tab:category}).

\subsubsection{Attitudes}
This category consisted of comments that offered straightforward positive or negative evaluations of Social.AI (see \autoref{tab:attitudes}). 
Users often emphasized novelty: ``\textit{love the idea, and love the logo and the design—in a world where Character.AI is so popular, this might be a hit as well.}'' This suggests that early adopters judged Social.AI's potential by situating it in comparison with existing platforms, highlighting perceived competitiveness in an already crowded market. Similarly, another comment framed the app as unexpectedly creative: ``\textit{That is creative! When I first read it, I was like, `…wtf,' though if you think deeply, many people really need this.}'' This indicated that the sense of surprise gradually gave way to recognition of possible social demand.
In contrast, negative voices dismissed the platform as redundant or harmful: ``\textit{This is spectacularly stupid. This is spectacularly dangerous.}'' Others dismissed the app as unnecessary duplication: ``\textit{Couldn’t I just open OpenAI or similar and write??}'', underscoring skepticism about whether multi-agent interaction adds genuine value.
This polarization illustrates the contested legitimacy of multi-agent social platforms at their early stage of public exposure, where optimism about innovation coexists with concerns about redundancy and toxicity.

\subsubsection{Motivations}
This category captured users’ reasons for engaging with Social.AI. Unlike the attitudes, these comments not only evaluated the platform but also revealed underlying drivers of participation. \x{Two} main motivational patterns emerged: \x{emotional} motivations and functional motivations.(see \autoref{tab:motivations}).

\paragraph{Emotional Motivations}
Unlike other one-on-one chatbots, Social.AI boasted millions of AI followers and instant replies. This uniqueness encouraged most users to anticipate novel experiences: ``\textit{Would be an interesting concept to see what life as a celeb is like. You make a post. Get 10000 upvotes in minutes.}'' 
Another experimented with drafts: ``\textit{Super interesting idea, I have been putting my tweet drafts in to see how the AIs react.}'' For others, emotional motives were more explicit. Some users explicitly contrasted Social.AI with disappointing or unwelcoming human interactions: ``\textit{Love this imaginary social network because the real ones have always let me down.}''  Users perceived Social.AI as more attentive and emotionally responsive than conventional social networks to reduce loneliness: ``\textit{I feel lonely on any other real people Twitter clone app because no one responds. But these bots feel pretty real to me.}'' 

\paragraph{Functional Motivations}
Users hoped that the extensive AI agents on the platform would expose them to diverse perspectives and broaden their thinking: ``\textit{As a playground to practice tweeting.}'' While others highlighted its utility in learning: ``\textit{I would like to use it for learning.}'' Users also used the platform for broadening their thinking: ``\textit{This is awesome. I could post any of my choices and get different perspectives on my choices, pros and cons, good and bad, then I get to decide how to move forward.}'' These cases suggested that, beyond entertainment, Social.AI is appropriated as an interactive resource for skill building and problem-solving.

\begin{table*}[t!]
\centering
\fontsize{8}{9}\selectfont
\begin{tabular}{lp{4cm}lp{4cm}}
\toprule
Code    & Description      &Frequency    & Example \\
\midrule
Social Empowerment     & The affirmation provided by AI feedback enhances users' social skills. & 13   &  I feel this app will make me feel more confident when sharing my ideas on the internet.  \\
Comfort      & Emotional reassurance provided by AI responses that alleviate loneliness or reduce stress. & 10   & This will be perfect as a first step in helping people with social anxiety.  \\
Being heard  & Users feel validated when their expressions, even if superficial, receive acknowledgment through replies.     & 7      & A place to share an idea and get tons of feedback about it.                   \\
Stimulation  & \x{Eliciting short-term excitement or reward responses.}    & 6   & An easy way to get feedback from multiple AI agents quickly.                    \\
\bottomrule
\end{tabular}
\caption{Table of codes, descriptions, frequencies, and examples within the \x{\textit{Perceived Social Support}} category.}
\label{tab:Mechanisms}
 \Description{Codes of \x{Perceived Social Support}. The table contains four columns: Code, Description, Frequency, and Example. The table contains four categories: Social Empowerment, Comfort, Being heard, and Stimulation. Readers can interact directly with the table to explore the specific coding definitions.}
\vspace{0em}
\end{table*}

\begin{table*}[t]
\centering
\fontsize{8}{9}\selectfont
\begin{tabular}{lp{4cm}lp{4cm}}
\toprule
Code    & Description    &Frequency    & Example \\
\midrule
Lack of Realism          & The content, timing, and length of agents' comments are not human-like enough. & 43                                     & I think I prefer it if the replies came in one at a time with a little delay. Getting a ton of instant replies is a bit overwhelming and breaks the immersion.  \\
Accessibility Issues     & Unable to use the application or access its core functionality, reporting a system failure that disrupted interaction.    & 40     & Why didn't I get a single reply on my post?                  \\
Feature Limitations      & Limitations of the platform's interactive features. & 19  & I would love to be able to upload photos into it.  \\
\bottomrule
\end{tabular}
\caption{Table of codes, descriptions, frequencies, and examples within the \textit{Technical Limitations} category.}
\label{tab:Limitations}
 \Description{Codes of Technical Limitations. The table contains four columns: Code, Description, Frequency, and Example. The table contains three categories: Lack of Realism, Accessibility Issues, and Feature Limitations. Readers can interact directly with the table to explore the specific coding definitions.}
\vspace{0em}
\end{table*}

\subsubsection{Perceived Social Support}
We identified four types of \x{perceived social support}, including social empowerment, comfort, being heard, and stimulation (see \autoref{tab:Mechanisms}).

\paragraph{Social Empowerment}
Some users described the platform as offering social empowerment, a low-stakes arena for testing ideas and building confidence. For example, one remarked: ``\textit{This app makes me feel more confident when sharing my ideas on the internet.}'' Another suggested its value for practice: ``\textit{Improved the idea and prepared for the real world.}'' The AI crowd was not simply a source of comfort or stimulation, but a rehearsal space for future engagement in human social environments. This indicated that users may see Social.AI not just as an emotional refuge, but as a training ground for navigating real-world interactions.

\paragraph{Comfort}
Beyond acknowledgment, some users stressed the comforting potential of AI responses. For individuals who had experienced social rejection or lacked supportive human networks, AI feedback was seen as a source of encouragement and affirmation: ``\textit{For someone who has been negged their whole life... they can actually start to build up some positive self-esteem, in a therapeutic way.}'' Another emphasized its therapeutic potential for vulnerable groups: ``\textit{This will be perfect as a first step in helping people with social anxiety.}'' Others highlighted that when encouragement is all one seeks, the identity of the responder matters little: ``\textit{If all they look for is support and encouragement, then the identity behind whoever interacts with them is not important.}''

\paragraph{Being heard}
Users valued having a safe outlet to share their thoughts, however trivial or unusual, and receive acknowledgment from multiple agents. As a user explained, ``\textit{This is actually a really cool idea. especially for when you want to just share random things and want to be heard.}'' Another drew an analogy to therapy: ``\textit{It’s like having a counselor who can listen to your crazy ass weird shit without blinking.}'' These remarks suggest that users were not necessarily seeking deep engagement, but rather the assurance that their expressions would not be ignored: ``\textit{Sometimes you just want to be heard.}'' The mere act of posting and receiving multiple comments, regardless of their quality of content, can create a psychological effect of being listened to.

\paragraph{Stimulation}
Some comments framed Social.AI as a form of stimulation, describing the platform as addictive or even pharmacological in its appeal. Users valued the rapid, multi-voice responses that offered immediate bursts of feedback: ``\textit{An easy way to get feedback from multiple AI agents quickly.}'' Others explicitly likened this to neurochemical reward: ``\textit{This is a drug to hit your dopamine receptors.}'' Some engaged in humorous exaggeration: ``\textit{an app where I can feel judged by millions of AI instead of just my friends.}''

\begin{table*}[t!]
\centering
\fontsize{8}{9}\selectfont
\begin{tabular}{lp{4cm}lp{4cm}}
\toprule
Code    & Description      &Frequency    & Example \\
\midrule
The Death of the Internet      & AI-generated content could drown out real human voices online, leading to a bot-dominated Internet.    & 86      & Humans are already and will furthermore be a minority on the internet.  \\
Illusory Sociality               & Social.AI creates a false sense of companionship that fails to capture the essence of social interaction. & 72      & The value of social network is the signal/noise ratio and the user base that makes it valuable.  \\
Ethical and Commercial Risks      & Concerns exist that Social.AI could harm humans or be used for commercial marketing.    & 37    & What if this algorithm is hacked and starts giving you the replies worse than imagination!  \\
Blurred Human–AI Boundaries      & Users struggle to distinguish between interacting with AI agents and real people.    & 28     & Are you a bot? Kidding…Or maybe I’m not.                   \\
Corrosion of Real-Life Social Interaction      & Reliance on AI-mediated interactions could weaken real-world social skills, erode interpersonal connections, and exacerbate loneliness. & 15     & AI has separated us and works to keep us isolated and controllable.   \\
\bottomrule
\end{tabular}
\caption{Table of codes, descriptions, frequencies, and examples within the \textit{Perceived Risks} category.}
\label{tab:Risks}
 \Description{Codes of Perceived Risks. The table contains four columns: Code, Description, Frequency, and Example. The table contains five categories: The Death of the Internet, Illusory Sociality, Ethical and Commercial Risks, Blurred Human–AI Boundaries, and Corrosion of Real-Life Social Interaction. Readers can interact directly with the table to explore the specific coding definitions.}
\vspace{0em}
\end{table*}

\subsubsection{Technical Limitations}
In contrast to discussions of emotional support, a large number of comments focused on the technical shortcomings of Social.AI, which directly constrained users' experience (see \autoref{tab:Limitations}).

\paragraph{Lack of Realism}
Users highlighted a lack of realism in the generated interactions. Replies sometimes contained fabricated facts: ``\textit{The replies often make up random facts and inaccurate bits or trivia.}'' Users also found the speed and uniformity of responses immersion-breaking: ``\textit{I think I prefer it if the replies came in one at a time with a little delay. Getting a ton of instant replies is a bit overwhelming and breaks the immersion.}'' Similarly, identical comment lengths reduced perceived authenticity: ``\textit{Change the length of the comments so that some AI leave longer comments and some are shorter.}''

\paragraph{Accessibility Issues}
Users first pointed to access barriers, noting that the app was unavailable on Android or the web: ``\textit{When roll out the Android app?}'' and ``\textit{Any plans to publish to the web? Would love access to a keyboard.}'' Others reported breakdowns in responsiveness, where posts failed to elicit any comments from the agents: ``\textit{I don't have any comments or anything else, what am I doing wrong?}'' These failures undermined the promise of instant feedback and left users questioning the reliability of the system.

\paragraph{Feature Limitations}
Users drew attention to feature limitations, requesting expansions that would make the platform feel closer to a real social media platform. Suggestions included uploading images and videos (``\textit{To update this app where you can upload videos and send images to everyone (ai).}''), displaying follower counts (``\textit{I think you guys should add the number of followers you can have.}''), enabling direct messages (''\textit{Please add dms.}''), or allowing AI agents to interact with each other (``\textit{Any chance the other users can start conversations with each other? Sometimes I just want to browse, not post!}''). Collectively, these critiques reveal how technical constraints not only hinder usability but also limit the plausibility of the social illusion that Social.AI seeks to generate.

\subsubsection{Perceived Risks}
Beyond technical flaws, users voiced wide-ranging fears about Social.AI, often framed through evocative metaphors and cultural references(see \autoref{tab:Risks}).

\paragraph{The Death of the Internet}
A particularly salient theme among users framed Social.AI as part of the broader trajectory of what they called the death of the Internet. The proliferation of bots represents not just a nuisance but an existential threat to the digital sphere, where human voices risk being drowned out by algorithmically generated content: ``\textit{I've recently given up all social media other than Reddit as it's just bots, AI content, mindless idiots, and political division.}'' The most pessimistic vision depicted humans as a dwindling minority: ``\textit{Humans are already and will furthermore be a minority on the internet, silently hiding, barely observing, for the uselessness of the content forbids attention and contributing to a sea of shitting bots it's useless.}''
These anxieties were further dramatized through cultural metaphors. The notion of ``Heaven banning'' imagined a future where human users are seamlessly displaced by AI-generated replicas of their friends. Users continue to interact online without realizing that authentic social ties have been replaced, producing an invisible expulsion from genuine human networks: ``\textit{For now. In another 5 years or so, it should be possible to create an AI model that uses a database of your friends' post history to accurately simulate their responses with a relatively high degree of accuracy. The issue would be if you ever saw them in real life. Though even that could be bypassed if your friends could see your posts and pictures and comment on them, to you, their interactions were hidden and replaced by AI content trained on their responses to simulate their personalities. Then there would be a seamless transition into the Heaven-banned state that wouldn't even be noticeable in the real world.}'' Another recurring theme was dystopian imagination: ``\textit{I can see people using this, but I also think that it is incredibly dystopian.'', with users likening Social.AI to ``a Black Mirror episode}'', and arguing that it was anti-society: ``\textit{Algo-fueled social media is literally the most powerful anti-social machine ever invented.}'' These comments framed Social.AI not merely as a flawed tool but as symptomatic of a trajectory where human agency and authenticity are displaced by algorithmic simulation.

\paragraph{Illusory Sociality}
Many doubted whether AI-driven exchanges could be considered genuine social interaction: ``\textit{As an art piece it’s great, as a social network it’s not social.}'' Because social interactions are formed by a group of real people, while AI isn't even human: ``\textit{Why call it social AI? It's not even social. Social comes from society, and society is a group of people. AI isn't people.}'' Others warned that Social.AI could devolve into a space of endless praise, eliminating dissent: ``\textit{Creating a literal echo chamber so you can talk to yourself and the computer.}'' Such dynamics erode the diversity and contestation that characterize real social life: ``\textit{Replaced by AI models that constantly agree and praise them, that would make anyone immediately suspicious. Whenever I go on any social media platform, I totally expect to be confronted with a certain amount of heated disagreements, vile insults, and outrageously stupid takes.}'' The notion of ``double falseness'' appeared: ``\textit{First social media took us away from real life, and now social AI is creating a fake social media, we're twice the amount of fakeness away from the actual reality. Maybe in the future, this will become more common than reality itself.}''

\paragraph{Ethical and Commercial Risks}
Concerns extended to mental health (``\textit{What if this algorithm is hacked and starts giving you the replies worse beyond imagination!}'' and ``\textit{Stop messing people's mental health!}''), privacy breaches (``\textit{Selling credible-looking profiles to scammers or startup influencers.}''), and covert commercialization (``\textit{Advertisers will be able to buy nudges that will get the AI users to mention certain products, share certain links.}''). Such fears echo long-standing critiques of algorithmic governance and capitalism, and are magnified by the multi-agent of Social.AI.

\paragraph{Blurred Human–AI Boundaries}
Comments revealed anxiety about not knowing who or what they were interacting with, and even gave rise to doubts about one's own identity: ``\textit{Are you a bot? Or maybe I am not. But honestly, it has become a legitimate question when interacting with any `person' online.}'' Some noted the hurt of being mistaken for bots themselves, underscoring how fragile identity has become in hybrid spaces: ``\textit{I have been accused of being a bot on my old account. It kind of hurt.}''

\paragraph{Corrosion of Real-Life Social Interaction}
Whereas digital bots were seen as hollowing out online communities, a deeper anxiety was that reliance on simulated interactions could gradually erode the fabric of human connection in the real world: ``\textit{This is incredibly stupid and beyond useless, actively corrosive to human connection and interaction.}'' In this narrative, Social.AI is not merely for virtual companionship but is subtly eroding humans' capacity to interact, express themselves, and communicate with others: ``\textit{Learning how to communicate ideas is a skill—and the recent infatuation with AI-generated art is threatening to kill it off, by de-incentivising people to do it.}'' The danger of this trend lies in its reinforcement of isolation: ``\textit{Nothing to further a feeling of alienation from other people like creating ways of never interacting with another person again.}''

\textbf{In summary}, our analysis of online discourse reveals diverse and ambivalent public attitudes toward Social.AI. Many users expressed interest in the potential to feel heard and empowered, reflecting predictions by CASA that users tend to treat AI as social actors, projecting onto them expectations of attention, care, and reciprocity. However, a larger portion of the public maintained a more detached view, regarding Social.AI primarily as a tool or a commercially driven platform, leading to prevalent concerns and resistance. This observation prompted us to question whether public discourse adequately represents actual users' lived experiences on such platforms, as online discourse is often shaped by influencers, reviewers, researchers, and critics, whose primary motives are evaluative or critical rather than personal or therapeutic.

%% file: Sections/04-study2.tex
\section{Study 2: Dairy Study}
\label{sec:diary}

To further explore \x{users' lived experiences on Social.AI—especially, whether they develop social expectations toward AI agents and how such expectations might fail—}we proceeded to the second stage: a 7-day diary study. This approach allows us to examine in depth how individuals experience and interpret their interactions with Social.AI over time.

\subsection{Participants}
We recruited participants from \x{a U.S. university and snowballing}. Recruitment notices were distributed via campus-wide emails, directing interested individuals to participate in a Qualtrics survey. This questionnaire collected demographic information from participants, prior familiarity with AI companion apps \x{(\eg Character.ai, Replika)}, and willingness to use the Social.AI application. \x{Twenty} participants were ultimately recruited and asked to use Social.AI daily for 7 consecutive days (see \autoref{tab:participants}).
\x{None of our diary-study participants were experienced Social.ai users before our research. This design choice was made partly because the platform is still relatively new, but more importantly, because we wanted to observe the complete trajectory from initial trial to sustained (or abandoned) use. Recruiting experienced users would have obscured these dynamics.}

\begin{table}[htbp]
\centering
\fontsize{8}{9}\selectfont
\begin{tabular}{lllp{3cm}}
\toprule
ID  &	Sex     &	Age             &	Familiarity with AI Companion Apps\\
\midrule
P1	&	Male	&	25-34 years old	&  Highly	\\
P2	&	Female	&	18-24 years old &  Moderately	\\
P3	&	Male	&	25-34 years old &  Slightly	\\
P4	&	Male	&	25-34 years old &  Slightly	\\
P5	&	Male	&	25-34 years old &  Moderately	\\
P6	&	Female	&	35-44 years old &  Slightly\\
P7	&	Female	&	18-24 years old &  Slightly\\
P8	&	Male	&	18-24 years old &  Moderately\\
P9	&	Female	&	18-24 years old &  Slightly\\
P10	&	Male	&	18-24 years old &  Slightly\\
P11	&	Male	&	18-24 years old &  Highly\\
P12	&	Female	&	25-34 years old &  Moderately	\\
P13	&	Female	&	18-24 years old &  Highly	\\
P14	&	Female	&	18-24 years old &  Highly	\\
P15	&	Female	&	18-24 years old &  Moderately	\\
P16	&	Female	&	18-24 years old &  Highly\\
P17	&	Male	&	18-24 years old &  Slightly\\
P18	&	Male	&	18-24 years old &	Moderately\\
P19	&	Male	&	18-24 years old &	Moderately\\
P20	&	Male	&	25-34 years old &	Highly\\ 
\bottomrule
\end{tabular}
\caption{\x{Information of the participants.}}
\label{tab:participants}
 \Description{Information of the 20 participants. The table contains 4 columns: ID, Sex, Age, and Familiarity with AI Companion Apps.  Readers can interact directly with the table to explore the specific coding definitions.}
\vspace{0em}
\end{table}

The journal prompts were modeled after Davis~\etal's study on adolescents' emotional regulation via Instagram~\cite{davis2025you}. Each participant recorded their interactions, emotional responses, and reflections daily, ultimately forming a rich diary text corpus. Each participant received a \$10 Amazon gift card as compensation.
Each morning, we emailed participants a link to that day's log, requesting they record: (i) duration of use; (ii) motivations and behaviors (\eg post, comment, and reply); (iii) emotional reactions; (iv) impact on daily life. This structure captured both immediate emotional fluctuations and deeper reflections on the role of AI-driven social interactions in everyday life.

\subsection{Data Analysis}
After collecting a 7-day diary from the participants, we conducted a thematic analysis to address RQ2 (\ie \x{What are the lived experiences of users during actual engagement with the multi-agent AI social platform?}) 
Two authors first familiarized ourselves with the data by repeatedly reading diary entries and highlighting significant moments of use. \x{With the research question in mind, the first author generated an initial set of open codes, such as \textit{Feeling Supported}, \textit{Perspective Diversity}, and \textit{Homogenized Responses}. Through two rounds of discussion, we refined and organized these codes into two main dimensions: (i) positive experiences on Social.AI, and (ii) negative experiences on Social.AI, leading to the final coding framework in \autoref{tab:S2 category}.} 
\x{The two coders then independently applied the code book to the full dataset. Inter-coder reliability reached a Cohen's Kappa of 0.89, indicating substantial agreement. Remaining discrepancies were resolved through discussion, and the final code book was consolidated through consensus.} These themes are elaborated with direct diary excerpts in the following section.

\subsection{Findings}

\x{As summarized in \autoref{tab:S2 category}, our participants reported both positive and negative aspects of socializing with Social.AI. Below, we report them in detail.}

\begin{table*}[htbp]
\centering
\fontsize{8}{9}\selectfont
\begin{tabular}{p{1.8cm}llp{6cm}}
\toprule
Category        & Code      &Frequency    & Example\\
\midrule
\multirow{7}{*}{Positive Aspects} & Feeling Supported     & 24 & They give pretty positive comments about everything, which is good. \\
 & Meaningful Responses   & 20  & Some of the advice from comments is good, and makes me think of my next steps. \\
 & Feeling Popular        & 17  & Got a bunch of replies instantly. \\
 & Perspective Diversity  & 14  &  Different fan communities have distinct perspectives and reference points when analyzing issues.\\
 \hline
\multirow{12}{*}{Negative Aspects} & Homogenized Responses & 80 &  I realized comments are pretty much using the same/similar sentences.\\
 & Shallow Emotional Connection & 41  &  The lack of human impact, or rather emotional impact from the bots, failed to impact my day. \\
 & Physical Tiredness               & 13  & I realize that there are too many replies. It is impossible to check them all. So I'm not motivated to check them frequently. \\
 & Social Pressure              & 7   &  Once a post received many responses, replying to even one would trigger even more, creating an endless cycle of conversational obligations \\
 & Ethical Responsibility       & 7   &   I felt bad for some reason that all these replies kept endlessly generating for my amusement.\\
\bottomrule
\end{tabular}
\caption{\x{Table of categories, codes, frequencies, and examples reported in the diaries.}}
\label{tab:S2 category}
 \Description{Codes of two categories. The table contains four columns: Category, Code, Frequency, and Example. The table contains two categories: Positive Aspects and Negative Aspects. Readers can interact directly with the table to explore the specific coding definitions.}
\vspace{0em}
\end{table*}

\subsubsection{Positive Aspects}

Participants reported \x{four main positive aspects} of using Social.AI: \x{feeling supported, receiving meaningful responses, feeling popular, and gaining perspective diversity}. 

\x{
\paragraph{Feeling Supported}
The most mentioned positive experience was that Social.AI's instant responses created a sense of social presence (N = 24).}
For example, P2 noted that ``\textit{the replies make me feel like I am not alone.}'' \x{The agents' encouraging feedback further reinforced this supportive experience.} 
P5 used the app under exam stress to ``\textit{complain}'', finding encouragement in the positive replies. P4 was surprised ``\textit{to see accounts replying to a post I made two days ago}'', \x{and experienced a sense of being constantly noticed. These experiences align with CASA that humans automatically perceive simple social cues, such as responses, attention, and sustained interaction, as genuine social presence, thereby triggering emotional responses.}

\paragraph{\x{Meaningful Responses}}
Some users engaged with Social.AI for more pragmatic purposes, such as seeking advice or real-life information (N = 20). P1 received restaurant recommendations and planned to ``\textit{check the restaurant it recommended}'' \x{, indicating that the agents' responses are practically executable.} When searching for exercise routines, P7 found that  ``\textit{the bots' responses were helpful and understood what I was trying to ask when seeking advice.}''
\x{When addressing more complex issues, agents' responses also prompted users to reflect. In discussion on job hunting and career choices, P10 noted that agents prompted him to consider job satisfaction, which he hadn’t thought to look at. Similarly, P2 said that ``\textit{some of the advice is good, and makes me think of my next steps.}'' 
}

\paragraph{Feeling Popular}
Unlike traditional social media, where users must build followers or one-on-one chatbots, Social.AI deploys a large number of agents. Once users post, many agents respond at once. This created a sense of instant visibility and a crowd-like atmosphere. Participants often felt surprised or even flattered (N = 17). For example, P9 wrote on the first day: ``\textit{Posting a comment and receiving instant replies.}'' Some participants even compared the experience to being an influencer. As P20 noted: ``\textit{When I post, so many AIs reply to me that I feel like a celebrity.}'' From the CASA perspective, users naturally equated the number of responses with popularity. The experience of being clustered is an important part of Social.AI's early appeal.

\paragraph{Perspective Diversity}
Social.AI deploys agents with features distinct personalities (\eg supporters, haters, pessimists), providing users with diverse perspectives (N = 14). P11 observed that ``\textit{different fan communities have distinct perspectives and reference points when analyzing issues.}'' P19 also enjoyed selecting different follower types, noting that ``\textit{it’s fun to see comments with different attitudes.}'' These moments created a perception that Social.AI could broaden one's thinking, giving users a taste of the diverse perspectives often found in large social crowds. As P16 put it, ``\textit{when AI sparks multiple viewpoints, it delivers a fresh and intriguing experience.}''

\subsubsection{Negative Aspects}

However, negative experiences were also extensive, reflecting a trajectory in which users' social expectations gradually collapsed. Participants reported dissatisfaction with the objects of interaction, including homogenized responses, shallow emotional connections, physical tiredness, social pressure, and ethical responsibility. These negative experiences represented a violation of CASA expectations, leading users to perceive the interaction as artificial, exhausting, or emotionally hollow.

\paragraph{Homogenized Responses}

Despite the appearance of a crowd, participants described homogenization across multiple dimensions, including speed, content, and interaction patterns, that made the environment feel artificial and a lack of social dynamics (N = 80). P17 noticed the responses were ``\textit{generated simultaneously}'', lacking the time lag in real social networks. Participants were also disappointed by the large amount of repetitive content: ``\textit{They often repeated in responses and were often verbatim from one another}.'' P2 attempted to diversify the responses by selecting different agent types, ``\textit{expecting something to be different}''. Yet she still received similar responses, and ``\textit{felt like it was a waste of time today}''. \x{P20 noted that in real groups, ``\textit{people exchanging and even debating viewpoints is important}'', yet on Social.AI, ``\textit{the agents only agree with me, and that makes me lose the desire to communicate.}'' There was not even a broader social space on the platform. P18 captured this clearly, ``\textit{every question I ask gets a barrage of AI responses, with no real interaction between them, which is a stark contrast to real life.}'' Without natural interaction between agents and observable group dynamics, users were trapped in a self-centered loop. P4 was expecting the app to ``\textit{have a main feed sort of feature}'', but did not find it. The engagement was ``\textit{entirely dependent upon my posting things and initiating conversation.}''} This structure was particularly disappointing for users who expected to behave as ``lurkers'', just observing organic group exchanges. As P10 explained, ``\textit{when I use social media in general, I am more of a lurker and only selectively participate in conversations myself.}'' However, Social.AI forced users into the center of activity, stripping away the passive participation modes common in real social platforms.

\paragraph{Shallow Emotional Connection}
Although the platform delivered positive replies, participants consistently noted that the emotional engagement felt scripted and ultimately hollow (N = 41). As P8 on day 2 noted: ``\textit{Comments that respond to me as if given a specified emotional prompt rather than an organic question}'', highlighting that the responses felt like canned empathy rather than authentic engagement. By day 5, he felt sad ``\textit{to talk to an empty set of bots}'' and ``\textit{lonelier than without using}''. Even when participants received positive responses, the effect was short-lived and quickly became hollow. P5 observed that ``\textit{maybe the positive vibe can help me deal with anxiety or stress, but not much and almost unnoticeable.}'' 
\x{The gap was especially noticeable when users sought deep empathy. P18 vented about the hurdles in his PhD application, hoping for emotional support right now, ``\textit{but most AI are just spouting grand theories}''. They argued that these agents cannot truly understand their questions in depth. P4 said that: ``\textit{I felt like many responses didn’t fully understand what I was talking about.}'' Similarly, P15 found ``\textit{responses vague and lacking in substance, rather than addressing specific contexts}''. These experiences collectively demonstrated the anticipated rupture in CASA: users instinctively perceived AI as an empathetic social being, yet when attempting to establish deeper emotional connections, they find that ``\textit{AI's emotional responses remain overly stable, making it hard to experience authentic human-like communication or emotional fluctuations during conversations}''. }

\paragraph{Physical Tiredness}
While the initial flood of responses made some participants feel popular, this stimulation quickly became exhausting (N = 13). The volume of replies produced by dozens of agents created a sense of cognitive overload that participants struggled to manage. P4 said: ``\textit{I was looking through the responses to the first post I made, which seemed to be endless. When you reached the end of the replies, it was more populated. I found it to be a bit unnerving.}'' P18 also found that ``\textit{a single post gets flooded with AI replies. The sheer volume of responses makes it impossible to follow every AI's exact message.}'' So he can only skim through them. P16 didn't have the patience to read all the replies, ``\textit{knowing they'll keep coming in batches.}'' From CASA's perspective, this marked another point of breakdown: instead of interpreting the responses as social cues, users began to perceive them as mechanical system output.

\paragraph{Social Pressure}
Beyond physical tiredness, participants reported experiencing social pressure (N = 7). Users feel socially accountable without receiving the affective rewards of real social interaction. P20 said that ``\textit{once a post received many responses, replying to one would trigger even more}'', creating an endless cycle of conversational obligations. Constant questioning and follow-up pressure make users feel monitored rather than accompanied. P6 described feeling pressured: ``\textit{many of the bots continue to check in, ask if I have found inspiration, tried this, or that… and I hate that. I interpret it as pressure.}'' For her, the platform not only pushed for updates but also distorted the meaning of participation. She admitted that ``\textit{their responses make me wonder if this tool is supposed to encourage me to do XYZ rather than be a place to meander. Even though they are gen AI, I feel pressure to think of something worth posting about–if I don't do that, they'll just keep talking about past posts.}'' Similarly, P7 was burdened by repeated prompts to update prior discussions: ``\textit{Many bots have now begun asking for follow-ups, and I haven’t felt motivated to reply to any of it, since I don’t know what to say about my day 1 conversations.}''

\paragraph{Ethical Responsibility}
Participants were aware that the system was designed entirely around them, with no real autonomy (N = 7). P8 realized the bots were ``\textit{given brief prompts about their personalities and how to respond.}'' P4 even felt guilty because countless AI replies were just for his entertainment: ``\textit{I felt bad for some reason that all these replies kept endlessly generating for my amusement.}'' \x{P20 articulated a similar discomfort, noting that the agents were controlled by algorithms and forced to respond to him, ``\textit{unable to express authentic opinions or form their own relationships}''.}

\textbf{In summary}, we found that users indeed apply human social rules to AI agents and project expectations from human interactions onto these systems. However, we also observed an apparent paradox: AI agents provided high-frequency, instant, and active feedback, yet users commonly reported feelings of emptiness, detachment, and even social decline after several days of use. Although most participants had experienced comfort, validation, or joy, these positive effects were fragile and short-lived. Through this empirical study, we also revealed distinct user needs and behaviors on multi-agent social platforms (discussed in detail in \autoref{ssec:discussion_uniqueness}).

%% file: Sections/05-discussion.tex
\section{Discussion}
\label{sec:discuss}

In this section, we synthesize findings from the previous sections and draw upon existing literature to discuss our major insights and their implications.

\subsection{Comparing Public Opinions with Users' Real Experiences}

This work presents two complementary studies: a public comment analysis and a diary study documenting users' real experiences with Social.AI. Both studies reveal a significant commonality: users' engagement with this AI-driven social space consistently reflects what Kefi et al.~\cite{kefi2024ai} termed a ``transitional zone'' within the CASA paradigm—a state where the technology is neither fully perceived as a mere tool nor completely accepted as a human. 
\xx{Across both studies, users who engaged with Social.AI used interpersonal norms and emotional responses, despite knowing it is an AI.}
\xx{Also, users often oscillated between tool- and social-oriented perceptions. For instance, a user might initially treat the AI as a tool but gradually develop social expectations, or begin with social expectations only to disengage and revert to a tool-oriented perspective following various ``break-in-immersion'' moments.}
These observations collectively suggest the inherent instability and fluidity of the CASA paradigm within AI-mediated environments.

However, the two studies also revealed notable divergences. Public comments more frequently expressed skepticism and pessimism regarding multi-agent social media. Concerns included the creation of echo chambers, the displacement of real-world social interaction, and the ``internet death'' dominated by AI. While our diary study identified some corroborating instances—such as participants expressing concerns about homogenized viewpoints—most users did not perceive these issues as severe. Since users were free to disengage from the platform at any time, the sense of threat remained minimal. Notably, some users even reported that interacting with multiple AI ultimately enhanced their motivation to engage in real-world social interactions.
Second, public discourse often focused on detached, objective risks like commercial manipulation and privacy breaches. In contrast, the diary study revealed that once engaged with the platform, users readily immersed themselves. Despite knowing they were interacting with AI agents, they still experienced authentic social experiences such as happiness and guilt. This demonstrates a clear disconnect between speculative public criticism and the complex experiences during actual use.

\subsection{\xx{AI-Dominant Social Media}}
\label{ssec:discussion_uniqueness}

\xx{Compared to the most common one-on-one chatbots (e.g., ChatGPT), the multi-agent AI social platform examined in this work features two key distinctions. First, it shifts from dyadic communication to group communication, moving beyond simple dialogue to exhibit characteristics of social media. Second, it transforms from a human-centric model to one where AI agents surround the human user, creating an AI-dominant environment. As such, we believe the most interesting aspect of this work lies in exploring \textbf{how human users experience and interact within this AI-dominant social media setting.}}

Naturally, some aspects of this experience align with findings from traditional chatbot research. For instance, the lack of deep understanding (as reported in \autoref{sec:diary}) is a limitation common to current chatbots. 
According to the social penetration theory, relationship depth depends on mutual self-disclosure and gradual trust-building~\cite{skjuve2021my}, no matter in dyads or groups~\cite{williams2010dyads}. 
Yet in both settings, AI often shows shallow depth and limited authenticity of emotion, failing to make users feel genuinely understood or cared for~\cite{wu2024social,hu2025ai}. 
Beyond this shared issue, however, the other identified empirical evidence reflects user needs that are more specific to AI-generated social media environments:

\textbf{Seeking Intergroup Diversity and Tension.}
The predominant theme of complaints we identified centers on the homogenization of AI agents. This suggests that a key appeal of multi-agent platforms lies in their potential to simulate collective intelligence, which is predicated on the dynamic interplay of diverse, and even conflicting, perspectives. As argued by Putnam's communication framework~\cite{putnam1988communication}, ``cognitive friction'' is a fundamental mechanism for group sense-making, where debate and disagreement are essential for forging durable understanding. In contrast, in our study, the systemic generation of homogeneous affirmation, even from personas assigned as critics, creates a paradox of artificial collectivity: the form of a group is present, but its core cognitive function is absent. Consequently, the multi-agent environment devolves into what users termed a ``one-man show''.


\textbf{Managing Collective Attention.}
Another user complaint involves the social pressure from multi-agent interactions. Compared to a dyadic chat, which facilitates a controlled, turn-based rhythm, the multi-agent environment creates a continuous, one-to-many stream of interaction, turning the user into a central node managing unsolicited inputs. This dynamic thus evokes \textit{evaluation apprehension}~\cite{cottrell1968social}, \ie an individual's concern about being observed and evaluated by others. 
By situating users under the persistent attention of AI agents, such platforms simulate a socially dense scenario wherein users must continually navigate their position as the focal point. Even when explicitly aware that responders are AI, and though the social pressure may differ from being surrounded by real people, users still experience a sense of being perpetually observed~\cite{siemon2023let}. For these users, the experience might be more overwhelming than empowering.

\textbf{Coping with Emergent Power.} 
The multi-agent environment uniquely gives rise to power dynamics that are structurally emergent rather than interpersonally negotiated~\cite{cialdini2004social,abrams2006social}. For example, in Social.AI, this manifests as a consistently distorted social hierarchy, where the user is positioned with absolute authority over a cohort of servile agents. The resulting ``dictator effect''—which a few participants found ethically disconcerting—illustrates how this artificial power structure disrupts the mutual reciprocity essential for meaningful connection, constituting a fundamental characteristic of such social spaces. While this structure may appeal to users who derive comfort from authoritative roles, it poses distinct psychological challenges for those troubled by its inherent power imbalance.

\subsection{Design Implications}
\x{Based on the above observations, we propose a set of design suggestions specific to future multi-agent social platforms.}

\textbf{Diversity in Agent Persona Design.}
Platforms should prevent all agents from saying the same flattering thing and balance the quality and quantity of responses. \x{This requires careful design of agents with divergent worldviews, values, communication styles, and their mutual relationships—not merely superficial role labels~\cite{niu2025scenario}. As seen in projects such as Cinema Multiverse Lounge~\cite{ryu2025cinema}, well-designed inter-worldview conflicts can foster engaging and reflective user experiences. Techniques such as Multi-Agent Debate~\cite{liang2024encouraging} could also be integrated to stimulate cognitive friction, thereby supporting genuine collective sense-making} and mitigating the echo-chamber effect.

\x{
\textbf{User-Configurable Social Attention \& Pacing.}
To mitigate social pressure, platforms could incorporate mechanisms for regulating social density and interaction tempo.} For example, limit the number of agents that can respond simultaneously or introduce natural delays. Drip-feed replies can make each reply more meaningful and avoid information overload. \x{As suggested by findings on group size effects in AI environments~\cite{song2025multi}, systems could also allow users to customize the number of active agents, set interaction rhythms (\eg turn-taking modes), or activate a ``low-attention'' mode to reduce the sense of being constantly observed.}

\x{
\textbf{Balanced Power \& Reciprocity by Design.}
Instead of reinforcing rigid user-as-master hierarchies, platforms could embed dynamic role rotation, agent-initiated negotiation, and mutual feedback mechanisms. Drawing on the Group Experience (GX) framework~\cite{lee2025beyond}, systems can mediate power dynamics explicitly—for instance, by making social roles transparent, amplifying minority agent viewpoints, and scaffolding more equitable forms of dialogue.

\textbf{Support for Pluralistic Social Goals.}
As users differ in their social comfort and power preferences (e.g., some seek authority, others collaboration), platforms should support multiple social modes. These could range from consensus-building discussions to structured debates, allowing users to select interaction paradigms that align with their psychological preferences and self-efficacy levels.
}

\xx{Collectively, our analysis demonstrates that AI-dominant social media operates on a logic distinct from that of traditional platforms, fundamentally challenging the human-centered paradigm that has long guided HCI and social computing. This signals a structural shift: \textbf{social experience is increasingly shaped by AI-driven architectures rather than human intention alone.} Consequently, our work calls for new theoretical and design frameworks capable of accounting for environments in which \textbf{AI functions not merely as a tool or an anthropomorphized actor, but as the dominant medium of sociality itself.}
This shift reframes central challenges for both design and research:
For design, the task moves beyond multiplying agents or perfecting anthropomorphism, toward actively engineering the architectural sociality of the platform itself, such as curating meaningful inter-agent diversity, enabling user-controllable social density, and calibrating ethical power relations. For research, it necessitates a reorientation from studying how users perceive AI, to analyzing how AI-dominant architectures actively shape human perception, social expectations, and ethical reasoning. More importantly, commercially available systems like Social.AI foreshadow a broader \textbf{agentic turn in social media}. As such AI-dominant social media platforms develop and proliferate, they surface profound questions for future research; for example, as the medium of sociality itself changes, will it erode human capacities for complex and dissenting engagement, or can it be designed to cultivate more resilient and equitable forms of digital life?}

\subsection{Limitations and Future Work}
This work has several limitations. 
First, our analysis focused exclusively on a single multi-agent platform, Social.AI. While this setting provides unique insights into the dynamics of AI-only social environments, the findings may not generalize to other platforms with hybrid human–AI participation or different design logics. Future studies could broaden the scope by comparing Social.AI with role-playing chatbots or mainstream social platforms that gradually integrate AI agents.
\x{Second, it should be noted that the attitudes visible in public comments (Study 1) may not reflect the actual experiences of users (Study 2). As discussed earlier, while some expectations and concerns expressed in the comments were indeed observed in the diary study, others were less apparent or even contradicted by actual user experiences.}
Additionally, our diary study had a relatively small sample, which limits the demographic diversity of our findings. Expanding the sample to include varied age groups, cultural backgrounds, and usage contexts could help reveal how different populations perceive the promises and pitfalls of multi-agent social interaction.
Moreover, longitudinal studies that follow users over extended periods, or field deployments where Social.AI is embedded into daily routines, would provide richer evidence about sustained emotional and social effects.


%% file: Sections/06-conclusion.tex
\section{Conclusion}

This study examined how users perceive and experience Social.AI, a multi-agent social platform where all interactions, except for their own, are generated by AI. We conducted two complementary studies: (i) a content analysis of 883 public comments and (ii) a 7-day diary study with \x{twenty} participants. \x{We identified both commonalities and divergences between public discourse and actual user experiences. While both studies reflect, to varying degrees, the CASA paradigm's claim that people anthropomorphize interactive technologies and project social expectations onto them, public discourse focused more on macro-level risks, while real users engaged in deeper, more active expectation-projection onto AI agents. Users' expectations were sometimes met but were frequently accompanied by patterns of disappointment and frustration, driven by homogenized responses, shallow emotional connections, physical tiredness, social pressure, and perceived ethical responsibility. Building on these findings, we discuss the distinctive features of multi-agent social platforms and provide design implications for future work.}

%% file: reference.bib
@book{abrams2006social,
  title={Social identifications: A social psychology of intergroup relations and group processes},
  author={Abrams, Dominic and Hogg, Michael A},
  year={2006},
  publisher = {Routledge},
  address = {London},
}

@misc{socialai,         
    title={SocialAI}, 
    author={Friendly Apps},
    howpublished={\url{https://socialai.co/}},
    year={2024},
    note={Last accessed: Sept 4, 2025}
}

@misc{introtosocialai,        
    title={Introduction to Social.AI in the Apple Store}, 
    author={Friendly Apps},
    howpublished={\url{https://apps.apple.com/us/app/socialai-ai-social-network/id6670229993}},
    year={2024},
    note={Last accessed: Sept 4, 2025}
}

@misc{LiangZiWeiZhiKuAI,
  title = {AI Companionship is Starting to Wane, with Leading Products Experiencing Nearly Zero Growth.},
  author = {Quantum Bit Think Tank},
  year={2025},
  howpublished = {\url{https://mp.weixin.qq.com/s/ettRVWL-dCaj0AV8qXnybw}},
  note ={Last accessed: Sept 4, 2025}
}

@inproceedings{xygkou2023conversation,
  title={The ``Conversation'' about Loss: Understanding How Chatbot Technology was Used in Supporting People in Grief},
  author={Xygkou, Anna and Siriaraya, Panote and Covaci, Alexandra and Prigerson, Holly Gwen and Neimeyer, Robert and Ang, Chee Siang and She, Wan-Jou},
  booktitle={Proceedings of the CHI conference on human factors in computing systems},
  pages={1--15},
  year={2023},
  organization={Association for Computing Machinery}
}

@inproceedings{xu2017new,
  title={A New Chatbot for Customer Service on Social Media},
  author={Xu, Anbang and Liu, Zhe and Guo, Yufan and Sinha, Vibha and Akkiraju, Rama},
  booktitle={Proceedings of the CHI Conference on Human Factors in Computing Systems},
  pages={3506--3510},
  year={2017},
  organization={Association for Computing Machinery}
}

@inproceedings{ryu2025cinema,
  title={Cinema Multiverse Lounge: Enhancing Film Appreciation via Multi-Agent Conversations},
  author={Ryu, Jeongwoo and Kim, Kyusik and Heo, Dongseok and Song, Hyungwoo and Oh, Changhoon and Suh, Bongwon},
  booktitle={Proceedings of the CHI Conference on Human Factors in Computing Systems},
  pages={1--22},
  year={2025},
  organization={Association for Computing Machinery}
}

@inproceedings{yu2024exploring,
  title={Exploring the impact of anthropomorphism in role-playing AI chatbots on media dependency: A case study of Xuanhe AI},
  author={Yu, Qiufang and Lan, Xingyu},
  booktitle={Proceedings of the International Symposium of Chinese CHI},
  pages={170--181},
  year={2024}
}

@inproceedings{gatti2013large,
  title={Large-Scale Multi-Agent-Based Modeling and Simulation of Microblogging-Based Online Social Network},
  author={Gatti, Ma{\'\i}ra and Cavalin, Paulo and Neto, Samuel Barbosa and Pinhanez, Claudio and dos Santos, C{\'\i}cero and Gribel, Daniel and Appel, Ana Paula},
  booktitle={International Workshop on Multi-Agent Systems and Agent-Based Simulation},
  pages={17--33},
  year={2013},
  organization={Springer}
}

@inproceedings{soriano2013multi,
  title={Multi-Agent Systems Platform For Mobile Robots Collision Avoidance},
  author={Soriano, Angel and Bernabeu, Enrique J and Valera, Angel and Vall{\'e}s, Marina},
  booktitle={International Conference on Practical Applications of Agents and Multi-Agent Systems},
  pages={320--323},
  year={2013},
  organization={Springer}
}

@inproceedings{dominguez2015scope,
  title={SCOPE: a Multi-Agent System Tool for Supply Chain Network Analysis},
  author={Dom{\'\i}nguez, Roberto and Cannella, Salvatore and Framinan, Jose M},
  booktitle={IEEE EUROCON International Conference on Computer as a Tool (EUROCON)},
  pages={1--5},
  year={2015},
  organization={IEEE}
}

@inproceedings{davis2025you,
  title={" You Go Through So Many Emotions Scrolling Through Instagram": How Teens Use Instagram To Regulate Their Emotions},
  author={Davis, Katie and Landesman, Rotem and Yoon, Jina and Kim, JaeWon and Munoz Lopez, Daniela E and Magis-Weinberg, Lucia and Hiniker, Alexis},
  booktitle={Proceedings of the CHI Conference on Human Factors in Computing Systems},
  pages={1--16},
  year={2025},
  organization={Association for Computing Machinery}
}

@inproceedings{nass1994computers,
  title={Computers Are Social Actors},
  author={Nass, Clifford and Steuer, Jonathan and Tauber, Ellen R},
  booktitle={Proceedings of the CHI Conference on Human Factors in Computing Systems},
  pages={72--78},
  year={1994},
  organization={Association for Computing Machinery}
}

@inproceedings{zhang2025dark,
  title={The Dark Side of AI Companionship: A Taxonomy of Harmful Algorithmic Behaviors in Human-AI Relationships},
  author={Zhang, Renwen and Li, Han and Meng, Han and Zhan, Jinyuan and Gan, Hongyuan and Lee, Yi-Chieh},
  booktitle={Proceedings of the CHI Conference on Human Factors in Computing Systems},
  pages={1--17},
  year={2025},
  organization={Association for Computing Machinery}
}

@inproceedings{chaves2018single,
  title={Single or Multiple Conversational Agents? An Interactional Coherence Comparison},
  author={Chaves, Ana Paula and Gerosa, Marco Aurelio},
  booktitle={Proceedings of the CHI Conference on Human Factors in Computing Systems},
  pages={1--13},
  year={2018},
  organization={Association for Computing Machinery}
}

@inproceedings{geng2025beyond,
  title={Beyond the Dialogue: Multi-Chatbot Group Motivational Interviewing for Premenstrual Syndrome (PMS) Management},
  author={Geng, Shixian and Inayoshi, Remi and Yang, Chi-Lan and Sramek, Zefan and Umeda, Yuya and Kasahara, Chiaki and Sato, Arissa J and Hosio, Simo and Yatani, Koji},
  booktitle={Proceedings of the CHI Conference on Human Factors in Computing Systems},
  pages={1--18},
  year={2025},
  organization={Association for Computing Machinery}
}

@inproceedings{gamage2025labeling,
  title={Labeling Synthetic Content: User Perceptions of Label Designs for AI-Generated Content on Social Media},
  author={Gamage, Dilrukshi and Sewwandi, Dilki and Zhang, Min and Bandara, Arosha K},
  booktitle={Proceedings of the CHI Conference on Human Factors in Computing Systems},
  pages={1--29},
  year={2025},
  organization={Association for Computing Machinery}
}

@inproceedings{tanprasert2024debate,
  title={Debate Chatbots to Facilitate Critical Thinking on YouTube: Social Identity and Conversational Style Make a Difference},
  author={Tanprasert, Thitaree and Fels, Sidney S and Sinnamon, Luanne and Yoon, Dongwook},
  booktitle={Proceedings of the CHI Conference on Human Factors in Computing Systems},
  pages={1--24},
  year={2024},
  organization={Association for Computing Machinery}
}

@inproceedings{afyouni2022living,
  title={Living one week with an autonomous Pepper in a rehabilitation center: lessons from the field},
  author={Afyouni, Alia and Ocnarescu, Ioana and Cossin, Isabelle and Kamoun, Emna and Mazel, Alexandre and Fattal, Charles},
  booktitle={IEEE International Conference on Robot and Human Interactive Communication},
  pages={554--559},
  year={2022},
  organization={IEEE}
}

@article{qiu2025exploring,
  title={Exploring Teenagers' Trust in Al Chatbots: An Empirical Study of Chinese Middle-School Students},
  author={Qiu, Siyu and Lin, Anqi and Wang, Shiya and Lan, Xingyu},
  journal={arXiv preprint arXiv:2512.06647},
  year={2025}
}

@inproceedings{maeda2024human,
  title={When human-AI interactions become parasocial: Agency and anthropomorphism in affective design},
  author={Maeda, Takuya and Quan-Haase, Anabel},
  booktitle={Proceedings of the ACM Conference on Fairness, Accountability, and Transparency},
  pages={1068--1077},
  year={2024},
  organization={Association for Computing Machinery}
}

@incollection{balaji2010introduction,
  title = {An Introduction to Multi-Agent Systems},
  booktitle = {Innovations in Multi-Agent Systems and Applications - 1},
  author = {Balaji, P. G. and Srinivasan, D.},
  year = {2010},
  pages = {1--27},
  publisher = {Springer},
  address = {Berlin, Heidelberg},
}

@article{walter2025artificial,
  title={Artificial Influencers and the Dead Internet Theory},
  author={Walter, Yoshija},
  journal={AI \& SOCIETY},
  volume={40},
  number={1},
  pages={239--240},
  year={2025},
  publisher={Springer},
}

@article{skjuve2021my,
  title={My Chatbot Companion-a Study of Human-Chatbot Relationships},
  author={Skjuve, Marita and F{\o}lstad, Asbj{\o}rn and Fostervold, Knut Inge and Brandtzaeg, Petter Bae},
  journal={International Journal of Human-Computer Studies},
  volume={149},
  pages={102601},
  year={2021},
  publisher={Elsevier},
  address = {}
}

@article{cialdini2004social,
  title={Social influence: Compliance and conformity},
  author={Cialdini, Robert B and Goldstein, Noah J},
  journal={Annual review of psychology},
  volume={55},
  number={1},
  pages={591--621},
  year={2004},
  publisher={Annual Reviews},
  address = {}
}

@article{li2021anthropomorphism,
  title={Anthropomorphism brings us closer: The mediating role of psychological distance in User-AI assistant interactions},
  author={Li, Xinge and Sung, Yongjun},
  journal={Computers in Human Behavior},
  volume={118},
  pages={106680},
  year={2021},
  publisher={Elsevier},
  address = {Netherland}
}

@article{niu2018anthropomorphizing,
  title={Anthropomorphizing information to enhance trust in autonomous vehicles},
  author={Niu, Dongfang and Terken, Jacques and Eggen, Berry},
  journal={Human Factors and Ergonomics in Manufacturing \& Service Industries},
  volume={28},
  number={6},
  pages={352--359},
  year={2018},
  publisher={Wiley Online Library},
  address = {}
}

@article{tajfel2001integrative,
  title={An integrative theory of intergroup conflict},
  author={Tajfel, Henri and Turner, John and Austin, William G and Worchel, Stephen},
  journal={Intergroup relations: Essential readings},
  pages={94--109},
  year={2001}
}

@article{herbener2025lonely,
  title={Are Lonely Youngsters Turning to Chatbots for Companionship? The Relationship Between Chatbot Usage and Social Connectedness in Danish High-school Students},
  author={Herbener, Arthur Bran and Damholdt, Malene Flensborg},
  journal={International Journal of Human-Computer Studies},
  volume={196},
  pages={103409},
  year={2025},
  publisher={Elsevier},
  address = {}
}

@article{siddals2024happened,
  title={“It Happened to be the Perfect Thing”: Experiences of Generative AI Chatbots for Mental Health},
  author={Siddals, Steven and Torous, John and Coxon, Astrid},
  journal={npj Mental Health Research},
  volume={3},
  number={1},
  pages={48},
  year={2024},
  publisher={Nature Publishing Group UK London}
}

@article{wu2024social,
  title={Social and ethical impact of emotional AI advancement: the rise of pseudo-intimacy relationships and challenges in human interactions},
  author={Wu, Jie},
  journal={Frontiers in psychology},
  volume={15},
  pages={1410462},
  year={2024},
  publisher={Frontiers Media SA}
}

@article{de2025ai,
  title={AI companions reduce loneliness},
  author={De Freitas, Julian and O{\u{g}}uz-U{\u{g}}uralp, Zeliha and U{\u{g}}uralp, Ahmet Kaan and Puntoni, Stefano},
  journal={Journal of Consumer Research},
  pages={ucaf040},
  year={2025},
  publisher={Oxford University Press}
}

@article{yin2024ai,
  title={AI can help people feel heard, but an AI label diminishes this impact},
  author={Yin, Yidan and Jia, Nan and Wakslak, Cheryl J},
  journal={Proceedings of the National Academy of Sciences},
  volume={121},
  number={14},
  pages={e2319112121},
  year={2024},
  publisher={National Academy of Sciences}
}

@article{chen2025mind,
  title={MIND: Towards Immersive Psychological Healing with Multi-agent Inner Dialogue},
  author={Chen, Yujia and Li, Changsong and Wang, Yiming and Xiao, Qingqing and Zhang, Nan and Kong, Zifan and Wang, Peng and Yan, Binyu},
  journal={arXiv preprint arXiv:2502.19860},
  year={2025}
}

@article{li2025dialogueagents,
  title={DialogueAgents: A Hybrid Agent-Based Speech Synthesis Framework for Multi-Party Dialogue},
  author={Li, Xiang and Pan, Duyi and Xiao, Hongru and Han, Jiale and Tang, Jing and Ma, Jiabao and Wang, Wei and Cheng, Bo},
  journal={arXiv preprint arXiv:2504.14482},
  year={2025}
}

@article{park2023choicemates,
  title={Choicemates: Supporting unfamiliar online decision-making with multi-agent conversational interactions},
  author={Park, Jeongeon and Min, Bryan and Son, Kihoon and Song, Jean Y and Ma, Xiaojuan and Kim, Juho},
  journal={arXiv preprint arXiv:2310.01331},
  year={2023}
}

@article{cemri2025multi,
  title={Why do multi-agent LLM systems fail?},
  author={Cemri, Mert and Pan, Melissa Z and Yang, Shuyi and Agrawal, Lakshya A and Chopra, Bhavya and Tiwari, Rishabh and Keutzer, Kurt and Parameswaran, Aditya and Klein, Dan and Ramchandran, Kannan and others},
  journal={arXiv preprint arXiv:2503.13657},
  year={2025}
}

@article{pauwels2025ai,
  title={AI-induced sexual harassment: Investigating Contextual Characteristics and User Reactions of Sexual Harassment by a Companion Chatbot},
  author={Pauwels, Harrison and Razi, Afsaneh and others},
  journal={arXiv preprint arXiv:2504.04299},
  year={2025}
}

@article{hu2025ai,
  title={AI as your ally: The effects of AI-assisted venting on negative affect and perceived social support},
  author={Hu, Meilan and Chua, Xavier Cheng Wee and Diong, Shu Fen and Kasturiratna, KTA Sandeeshwara and Majeed, Nadyanna M and Hartanto, Andree},
  journal={Applied Psychology: Health and Well-Being},
  volume={17},
  number={1},
  pages={e12621},
  year={2025},
  publisher={Wiley Online Library},
  address = {}
}

@article{dorri2018multi,
  title={Multi-agent systems: A survey},
  author={Dorri, Ali and Kanhere, Salil S and Jurdak, Raja},
  journal={Ieee Access},
  volume={6},
  pages={28573--28593},
  year={2018},
  publisher={IEEE},
  address = {}
}

@article{cai2016general,
  title={A general multi-agent control approach for building energy system optimization},
  author={Cai, Jie and Kim, Donghun and Jaramillo, Rita and Braun, James E and Hu, Jianghai},
  journal={Energy and Buildings},
  volume={127},
  pages={337--351},
  year={2016},
  publisher={Elsevier},
  address = {}
}

@article{nguyen2012agent,
  title={Agent based restoration with distributed energy storage support in smart grids},
  author={Nguyen, Cuong P and Flueck, Alexander J},
  journal={IEEE Transactions on Smart Grid},
  volume={3},
  number={2},
  pages={1029--1038},
  year={2012},
  publisher={IEEE},
  address = {}
}

@article{milli2025engagement,
  title={Engagement, user satisfaction, and the amplification of divisive content on social media},
  author={Milli, Smitha and Carroll, Micah and Wang, Yike and Pandey, Sashrika and Zhao, Sebastian and Dragan, Anca D},
  journal={PNAS nexus},
  volume={4},
  number={3},
  pages={pgaf062},
  year={2025},
  publisher={Oxford University Press US}
}

@article{metzler2024social,
  title={Social drivers and algorithmic mechanisms on digital media},
  author={Metzler, Hannah and Garcia, David},
  journal={Perspectives on Psychological Science},
  volume={19},
  number={5},
  pages={735--748},
  year={2024},
  publisher={Sage Publications Sage CA: Los Angeles, CA}
}

@article{williams2010dyads,
  title={Dyads can be groups (and often are)},
  author={Williams, Kipling D},
  journal={Small Group Research},
  volume={41},
  number={2},
  pages={268--274},
  year={2010},
  publisher={Sage Publications Sage CA: Los Angeles, CA}
}

@article{ng2025dual,
  title={The Dual Personas of Social Media Bots},
  author={Ng, Lynnette Hui Xian and Carley, Kathleen M},
  journal={arXiv preprint arXiv:2504.12498},
  year={2025}
}

@article{ng2025global,
  title={A global comparison of social media bot and human characteristics},
  author={Ng, Lynnette Hui Xian and Carley, Kathleen M},
  journal={Scientific Reports},
  volume={15},
  number={1},
  pages={10973},
  year={2025},
  publisher={Nature Publishing Group UK London}
}

@article{schmuck2020perceived,
  title={Perceived threats from social bots: The media's role in supporting literacy},
  author={Schmuck, Desir{\'e}e and Von Sikorski, Christian},
  journal={Computers in Human Behavior},
  volume={113},
  pages={106507},
  year={2020},
  publisher={Elsevier}
}

@article{omeish2024investigating,
  title={Investigating the impact of AI on improving customer experience through social media marketing: An analysis of Jordanian Millennials},
  author={Omeish, Fandi and Al Khasawneh, Mohammad and Khair, Nadine},
  journal={Computers in human behavior reports},
  volume={15},
  pages={100464},
  year={2024},
  publisher={Elsevier}
}

@article{shao2018spread,
  title = {The Spread of Low-Credibility Content by Social Bots},
  author = {Shao, Chengcheng and Ciampaglia, Giovanni Luca and Varol, Onur and Yang, Kai-Cheng and Flammini, Alessandro and Menczer, Filippo},
  year = {2018},
  journal = {Nature Communications},
  volume = {9},
  number = {1},
  pages = {4787},
  publisher = {Nature Publishing Group}
}

@article{luo2023rise,
  title={Rise of social bots: The impact of social bots on public opinion dynamics in public health emergencies from an information ecology perspective},
  author={Luo, Han and Meng, Xiao and Zhao, Yifei and Cai, Meng},
  journal={Telematics and Informatics},
  volume={85},
  pages={102051},
  year={2023},
  publisher={Elsevier}
}

@article{song2025multi,
  title={Multi-agents are social groups: Investigating social influence of multiple agents in human-agent interactions},
  author={Song, Tianqi and Tan, Yugin and Zhu, Zicheng and Feng, Yibin and Lee, Yi-Chieh},
  journal={Proceedings of the ACM on Human-Computer Interaction},
  volume={9},
  number={7},
  pages={1--33},
  year={2025},
  publisher={ACM New York, NY, USA}
}

@article{siemon2023let,
  title={Let the computer evaluate your idea: evaluation apprehension in human-computer collaboration},
  author={Siemon, Dominik},
  journal={Behaviour \& Information Technology},
  volume={42},
  number={5},
  pages={459--477},
  year={2023},
  publisher={Taylor \& Francis},
  address = {}
}

@article{putnam1988communication,
  title={Communication, conflict, and dispute resolution: The study of interaction and the development of conflict theory},
  author={Putnam, Linda L and Folger, Joseph P},
  journal={Communication Research},
  volume={15},
  number={4},
  pages={349--359},
  year={1988},
  publisher={Sage Newbury Park},
  address = {}
}

@article{jiang2023communitybots,
  title={CommunityBots: creating and evaluating A multi-agent chatbot platform for public input elicitation},
  author={Jiang, Zhiqiu and Rashik, Mashrur and Panchal, Kunjal and Jasim, Mahmood and Sarvghad, Ali and Riahi, Pari and DeWitt, Erica and Thurber, Fey and Mahyar, Narges},
  journal={Proceedings of the ACM on Human-Computer Interaction},
  volume={7},
  number={CSCW1},
  pages={1--32},
  year={2023},
  publisher={ACM New York, NY, USA}
}

@article{chaves2021should,
  title={How should my chatbot interact? A survey on social characteristics in human--chatbot interaction design},
  author={Chaves, Ana Paula and Gerosa, Marco Aurelio},
  journal={International Journal of Human--Computer Interaction},
  volume={37},
  number={8},
  pages={729--758},
  year={2021},
  publisher={Taylor \& Francis},
  address = {Philadelphia}
}

@article{chen2024effects,
  title={Effects of anthropomorphic design cues of chatbots on users’ perception and visual behaviors},
  author={Chen, Jiahao and Guo, Fu and Ren, Zenggen and Li, Mingming and Ham, Jaap},
  journal={International journal of human--computer interaction},
  volume={40},
  number={14},
  pages={3636--3654},
  year={2024},
  publisher={Taylor \& Francis},
  address = {Philadelphia}
}

@article{kefi2024ai,
  title={AI-enabled social support chatbot usage: flowing ambivalence and liminalities},
  author={Kefi, Hajer and Khelladi, Insaf and Mani, Zied and Veg-Sala, Nathalie},
  journal={Journal of Decision Systems},
  pages={1--24},
  year={2024},
  publisher={Taylor \& Francis},
  address = {Philadelphia}
}

@article{ferrara2016rise,
  title={The rise of social bots},
  author={Ferrara, Emilio and Varol, Onur and Davis, Clayton and Menczer, Filippo and Flammini, Alessandro},
  journal={Communications of the ACM},
  volume={59},
  number={7},
  pages={96--104},
  year={2016},
  publisher={ACM New York, NY, USA}
}

@article{reeves1996media,
  title={The media equation: How people treat computers, television, and new media like real people},
  author={Reeves, Byron and Nass, Clifford},
  journal={Cambridge, UK},
  volume={10},
  number={10},
  pages={19--36},
  year={1996},
  publisher={},
  address = {}
}

@article{li2023you,
  title={Are you in a masquerade? exploring the behavior and impact of large language model driven social bots in online social networks},
  author={Li, Siyu and Yang, Jin and Zhao, Kui},
  journal={arXiv preprint arXiv:2307.10337},
  year={2023}
}

@article{gambino2020building,
  title={Building a stronger CASA: Extending the computers are social actors paradigm},
  author={Gambino, Andrew and Fox, Jesse and Ratan, Rabindra A},
  journal={Human-Machine Communication},
  volume={1},
  pages={71--85},
  year={2020},
  publisher={Communication and Social Robotics Labs Kalamazoo, Michigan}
}

@inproceedings{park2023generative,
  title={Generative agents: Interactive simulacra of human behavior},
  author={Park, Joon Sung and O'Brien, Joseph and Cai, Carrie Jun and Morris, Meredith Ringel and Liang, Percy and Bernstein, Michael S},
  booktitle={Proceedings of the annual acm symposium on user interface software and technology},
  pages={1--22},
  year={2023},
  organization={Association for Computing Machinery}
}

@article{cottrell1968social,
  title={Social facilitation of dominant responses by the presence of an audience and the mere presence of others.},
  author={Cottrell, Nickolas B and Wack, Dennis L and Sekerak, Gary J and Rittle, Robert H},
  journal={Journal of personality and social psychology},
  volume={9},
  number={3},
  pages={245},
  year={1968},
  publisher={American Psychological Association},
  address = {}
}

@inproceedings{liang2024encouraging,
  title={Encouraging divergent thinking in large language models through multi-agent debate},
  author={Liang, Tian and He, Zhiwei and Jiao, Wenxiang and Wang, Xing and Wang, Yan and Wang, Rui and Yang, Yujiu and Shi, Shuming and Tu, Zhaopeng},
  booktitle={Proceedings of the Conference on Empirical Methods in Natural Language Processing},
  pages={17889--17904},
  year={2024},
  organization={Association for Computational Linguistics}
}

@inproceedings{lee2025beyond,
  title={Beyond Individual UX: Defining Group Experience (GX) as a New Paradigm for Group-centered AI},
  author={Lee, Soohwan and Hwang, Seoyeong and Lee, Kyungho},
  booktitle={Companion Publication of the ACM Designing Interactive Systems Conference},
  pages={357--362},
  year={2025},
  organization={Association for Computing Machinery}
}

@inproceedings{niu2025scenario,
  title={Scenario, Role, and Persona: A Scoping Review of Design Strategies for Socially Intelligent AI Agents},
  author={Niu, Ruowen and Hu, Jiaxiong and Peng, Siyu and Cao, Caleb Chen and Liu, Chengzhong and Han, Sirui and Guo, Yike},
  booktitle={Proceedings of the Extended Abstracts of the CHI Conference on Human Factors in Computing Systems},
  pages={1--9},
  year={2025},
  organization={Association for Computing Machinery}
}
